\documentclass[prb,aps,showpacs,showkeys]{revtex4}

\usepackage{graphicx}

\begin{document}

\title{Electron--phonon interaction effects in semiconductor quantum dots:
a non-perturbative approach}
\author{M I Vasilevskiy}
\affiliation{Centro de F\'{\i}sica, Universidade do Minho, 4710-057 Braga, Portugal}
\email{mikhail@fisica.uminho.pt}
\author{E V Anda}
\affiliation{Departamento de F\'{\i}sica, Pontif\'{\i}cia Universidade Cat\'olica,
Rio de Janeiro-RJ, Brazil}
\email{anda@fis.puc-rio.br}
\author{S S Makler}
\affiliation{Instituto de F\'{\i}sica, Universidade Federal Fluminense,
Niter\'oi, RJ, Brazil}
\affiliation{Instituto de F\'{\i}sica, Universidade do Estado do Rio de Janeiro,
Rio de Janeiro-RJ, Brazil}
\email{sergio@if.uff.br}
\date{\today}

\begin{abstract}
Multiphonon processes in a model quantum dot (QD) containing two electronic states
and several optical phonon modes are considered taking into account both intra- and
inter-level terms. The Hamiltonian is exactly
diagonalized including a finite number of multi-phonon processes,
large enough as to guarantee that the result can be considered
exact in the physically important region of energies. The physical properties are
studied by calculating the
electronic Green's function and the QD dielectric function.
When both the intra- and inter-level interactions are included, the calculated
spectra allow for explanation of several previously published experimental results
obtained for spherical and self-assembled QDs, such as enhanced 2LO phonon replica
in absorption spectra and up-converted
photoluminescence. An explicit calculation of the spectral
line shape due to intra-level interaction with a continuum of acoustic phonons is
presented, where the multi-phonon processes also are shown to be important. It is pointed out
that such an interaction, under certain conditions, can lead to relaxation in the otherwise
stationary polaron system.
\end{abstract}


\maketitle
\newpage

\section{Introduction}
The important role of phonons influencing the properties of systems based on semiconductor quantum dots (QDs) has been
 demonstrated in many works reviewed in Refs.\cite{Banyai,Heitz02}. The spatial confinement effect on optical phonons
in QDs has been studied
experimentally and understood theoretically
\cite{Krauss,Chamberlain,Vasilevskiy,Vasilevskiy1,Pokatilov}. Acoustic phonons have also
received some attention in connection with the low-frequency Raman scattering
in spherical\cite{Verma,Savoit} and self-assembled\cite{Tenne,Cazayous2000} QDs,
 and homogeneous broadening of the spectral lines \cite{Palinginis,Warming,Krummheuer}.
Electron-phonon interaction in QDs remains, however, a
controversial subject. While calculations performed for II-VI and
III-V dots generally agree that the exciton-phonon (e-ph) coupling
strength is reduced in nanocrystals compared to the bulk, they
disagree regarding the numbers and trends in scaling of this interaction with QD
size. (Electron-hole pairs will be traditionally called here excitons
independently of the importance of the Coulomb interaction.) The calculated values vary substantially depending
on the approximations used, but the Huang-Rhys parameter for the
lowest exciton state usually does not exceed $S\approx 0.1-0.2$ for II-VI for spherical QDs\cite{HR1,HR2} and is
probably an order of magnitude smaller for III-V self-assembled dots\cite{Heitz99}.

Turning to the experimental data, the exciton-optical-phonon
coupling strength is most frequently obtained from PL spectra.
Usually there is a large Stokes shift of the excitonic PL band
with respect to the absorption in small II-VI QDs
\cite{Banyai,Bawendi}, which can be explained by strong
exciton-phonon coupling (Huang-Rhys parameter of the order of 1\cite{Juyckel,Bissiri}).
However, as mentioned above, the calculated values happen to be one or two orders of
magnitude smaller. Phonon replica in the
PL spectra caused by the recombination of excitons in QDs were
found to disagree with the well-known Franck-Condon progression both in their spectral positions\cite{Bissiri,Ignatiev}
 and relative intensities\cite{Bissiri,Fomin,Rho}. Although both the strong Stokes shift and
apparently large intensity of the first phonon satellite
(relative to the zero phonon line) can be characteristic of
the QD size distribution and not of a single dot \cite{Bawendi},
these results indicate that the e-ph interaction in QDs must be
considered carefully. Further evidence for this comes from the
optical absorption measurements. LO-phonon related
features were observed in absorption spectra of InAs/GaAs QDs and 2LO
\cite{Fafard,Schmidt,Lemaitre,Toda} and even 3LO\cite{Heitz} phonon replica were found extraordinarily strong
 compared to the 1LO phonon satellite. At least some of these works were performed using single QD spectroscopy, so,
ensemble effects can probably be ruled out in this case. Since no phonon replica were found in the corresponding
 emission spectra, there is no way to explain these results in terms of the Franck-Condon progression with any value of $S$.
The temperature dependence of the
homogeneous broadening of the absorption spectra of spherical II-VI QDs studied in Refs.\cite{HR1,Rakovich}
was found to have a clear contribution of optical phonons. This is unexpected, because coupling of optical phonons
 to a single electron or exciton level should lead to the appearance of discrete satellites and not to the level broadening.
Thinking in terms
of Fermi's Golden Rule, the broadening could be a lifetime effect owing to the electron (or exciton) transition to another
 state with emission or absorption of an optical phonon. However, it would
require strict energy conservation in the electron-phonon
scattering, i.e. exact resonance between the optical phonon energy
and level spacing, which should be rather accidental. This
kind of argument also justified the theoretical concept of ``phonon
bottleneck'', a very slow carrier relaxation which should be inherent to small
QDs\cite{Heitz02}. Nevertheless, an
efficient phonon-mediated carrier relaxation has been
reported in a number of works \cite{Ignatiev,Heitz,Marcinkevicius}. All
these experimental results imply that multiphonon processes are
important and that the e-ph interaction in QDs must be treated in a
non-perturbative way, even for the moderate values of the coupling constants
 coming out from the calculations\cite{Hameau}.
An important ingredient to be included is the non-adiabaticity of this
interaction \cite{Heitz02,Pokatilov,Fomin} leading to a phonon-mediated coupling of
different electronic levels, even if they are separated by an energy quite different
from the optical phonon energy. This is essential for understanding those experimental results which are
 in clear disagreement with the single level-generated Franck-Condon progression.

Polaron effects in QDs have been studied theoretically in
several recent papers \cite{Inoshita,Kral,Stauber,Jasak}. The model
considered in these works included two electronic levels and several
Einstein phonon modes. The one-electron spectral function was
obtained applying either self-consistent perturbation theory
approximations \cite{Inoshita,Kral} or exactly, using a combined
analytical and numerical approach \cite{Stauber}. The results
calculated using the perturbation theory approaches show shift and
broadening of the levels, even for a sufficiently large detuning
(defined as $\Delta^\prime  = \Delta -\hbar \omega _0$ where  $\Delta =(\varepsilon _2  -
\varepsilon _1)$, $\varepsilon _1$  and  $\varepsilon _2$ are
the electron energy levels and $\hbar \omega _0$ is the phonon
energy). However, as pointed out in Refs.\cite{Stauber,Levetas},
this broadening is an artifact. The exact spectral function should
consist of  $\delta$-functions (if no further effects are involved).
 Moreover, even if these
$\delta$-functions are broadened artificially, the self-consistent
perturbation theory approximations are not able to reproduce the
structure of the spectrum calculated exactly using the approach
proposed in Ref.\cite{Stauber}. The method of Ref.\cite{Stauber}
based on the Gram-Schmidt orthogonalization procedure is, however,
limited to the case when all the optical phonons have the same energy.
In reality, confined optical phonon modes are characterized by
different frequencies within the band of the
corresponding bulk material. Usually there are a small number of
such modes that interact more intensively with electrons or
excitons \cite{Chamberlain,Vasilevskiy1,Vasilevskiy2}. The
incorporation of this multiplicity seems to be important for
comparison of calculated and experimental results. The limitation of the approach\cite{Jasak}
 (based on the Davydov canonical transformation) consists in neglecting the multiphonon processes.

Another important issue is the role of acoustic phonons. Even though some acoustic modes can be confined to the
QD and therefore have discrete energies, there should be a continuum of modes within a certain energy range.
 The interaction with this continuum should smooth out the polaron spectrum (otherwise consisting of $\delta-$functions,
 as noted above). The lineshape is not expected to be a simple Lorentzian for a QD, instead, acoustic phonon sidebands
are formed for a discrete electronic level \cite{Krummheuer,Zimmermann2}. Considering the simultaneous interaction of
confined excitons with acoustic and optical phonons can reveal some new effects and,
 to the best of our knowledge, has not been performed beyond the one-level model \cite{SR}.

In this paper, we propose a non-perturbative approach to the
calculation of the phonon effects on the electron spectral
function, and emission and absorption spectra of a QD, based on
direct numerical diagonalization of the Hamiltonian matrix
including a small number of electronic levels and several optical
phonon modes of different energy. Such a straightforward method was
 employed in Ref.\cite{Savona} for phonons strongly coupled to an electron on
 a deep donor center.
This approach is developed further here so as to allow for the
incorporation of the interaction with (virtually all) acoustic
phonon modes, under the condition that the inter-level coupling
mediated by solely acoustic phonons can be neglected. We present
calculated results which elucidate the influence of various
relevant parameters, such as inter-level spacing, coupling strengths and
temperature, on the optical spectra of a model QD.
The paper is organized as follows. In Sec. II, we introduce the
Hamiltonian matrix to be diagonalized in order to obtain the
polaron states, and define the electron Green's function. In
Sec. III, we derive the expression for the imaginary part of the
exciton dielectric function and present the absorption and
emission spectra calculated for different values of the relevant parameters.
 Sec. IV is devoted to the interaction with
acoustic phonons and its incorporation into the calculation of the
polaron spectra. We discuss the calculated results, make comparison with experimental data, and conclude in
Sec. V.

\section{Electron--optical-phonon interaction: the non-pertur\-ba\-tive
solution}
Our model system consists of two electronic levels, both coupled
to {\it N} phonon modes (of frequencies $\omega _\nu$). It is described by the Hamiltonian,
\begin{equation}
H_1 = \sum\limits_{i = 1}^2 {\varepsilon _i a_i^ +  a_i^{ } }  + \sum
\limits_{\nu  = 1}^N {\hbar \omega _\nu  } {b_\nu ^ +  b_\nu ^{ }
} + \sum \limits_{ij}^2 {\sum \limits_{\nu  = 1}^N {g
_{ij}^{\scriptstyle  \hfill \atop  \scriptstyle \nu  \hfill} }
a_i^ +  a_j \left( {b_\nu ^ +
 + b_\nu ^{ } } \right)}
\label{Ham1}
\end{equation}
where $a_i^+ ,a_i^{ } $  are the fermion creation and annihilation
operators for electrons (or holes) and $b_\nu ^ + , b_\nu^{ } $ are the operators for
phonons.
The interaction with optical phonons occurs predominantly
through the Fr\"ohlich-type mechanism and the Hamiltonian matrix
elements between exciton states $i$ and $j$ are given by
\cite{Chamberlain,Vasilevskiy1},
\begin{equation}
g _{ij}^{\nu}  = e\int {\Psi _i^* (\vec r_e } ,\vec r_h
)\left[ {\phi _{\nu} (\vec r_h ) - \phi _{\nu} (\vec r_e )}
\right]\Psi _j (\vec r_e ,\vec r_h )d\vec r_e d\vec r_h
\label{alpha}
\end{equation}
where $\phi _{\nu}$ is the electrostatic potential created by the
$\nu$-th phonon mode and $\Psi _i$ the exciton wavefunction of state
$i$. In the following, we shall also use dimensionless interaction constants
$\alpha _{ij}^{\nu}=g _{ij}^{\nu}/(\hbar\omega _\nu)$ and omit the superscript when only one phonon mode is considered.

If $g _{12}^\nu $ were equal to zero, the
Hamiltonian (\ref{Ham1}) would be exactly solvable, because there would be
no interference between two electronic levels. The one-level model (known as independent boson model)
has analytical solution \cite{Mahan}. The polaron
spectrum, in the case of a single phonon mode, consists of equidistant peaks
 (separated by the phonon energy $\hbar
\omega _0$). This is where the Franck-Condon progression comes from.

The case of $\alpha _{12}  \ne 0$, $\alpha _{11}  =
\alpha _{22}  = 0$ for a single phonon mode, in the so called rotation wave (RW)
approximation \cite{QuantumOptics} (which consists in neglecting the terms
$a_2^+ a_1^{ } b_0 ^+$ and $a_2^{ }a_1^+ b_0 ^{ }$ in
(\ref{Ham1})), also allows for analytical solution, with the
spectrum given by
\begin{equation}
E_{1,2} (m) = ({\Delta^\prime    \mathord{\left/ {\vphantom {\Delta^\prime    2}}
\right. \kern-\nulldelimiterspace} 2}) \pm \sqrt {{{({\Delta^\prime }  ^2 }
\mathord{\left/ {\vphantom {{({\Delta^\prime  } ^2 } 4}} \right.
\kern-\nulldelimiterspace} 4}) + (m + 1)(g _{12}^{})^2 }
 + (m+1)\hbar\omega _0 \label{branches}
\end{equation}
where $m=0,1,2,\dots $ is the number of phonons in the mixed state. In addition to (\ref{branches}), there is
a state with $E=0$.

The general case, which is of interest here, can be treated by
mapping the many-body problem onto a single particle problem in a
Fock space of higher dimension \cite{Bonca,Makler}. It is natural
to consider a basis $\left| {n_1 n_2 \left\{ {m_\nu  } \right\}}
\right\rangle$   where  $n_r=0,1$ is the number of fermions on
level $r$ and $m_{\nu}$ is the number of phonons of mode $\nu$. In
principle, the Hamiltonian matrix is infinite but, as it will be
shown below, one can truncate the Fock space by allowing a certain
maximum number of phonons for each mode and obtain a very accurate
solution. Since we are interested in the case when there is a
single fermion in the dot, it is necessary to consider only the
states $\left| {10\left\{ {m_\nu  } \right\}} \right\rangle$  and
$\left| {01\left\{ {m_\nu  } \right\}} \right\rangle$. The
necessary matrix elements are:
\[
H_{ij}=\left[ {\varepsilon _i \delta _{ij} + \sum\limits_{\nu  =
1}^N {m_\nu \hbar \omega _\nu  } } \right] \prod\limits_\nu
{\delta _{m_\nu ^{'} m_\nu } }  +\sum\limits_{\nu = 1}^N {g
_{ij}^\nu \left[ {\sqrt {m_\nu   + 1} \delta _{m_\nu ^{'} m_\nu +
1}}  +  \sqrt {m_\nu  }\delta _{m_\nu ^{'} m_\nu   - 1} \right]
\prod\limits_{\mu  \ne \nu }{\delta _{m_\mu ^{'} m_\mu } }
}\nonumber
\]
where $i$ and $j$ abbreviate the basis vectors  $\left| {10\left\{
{m_\nu  } \right\}} \right\rangle$ ($i=1$) and  $\left| {01\left\{
{m_\nu  } \right\}} \right\rangle$ ($i=2$). The dimension of the
matrix is $2m_1 \cdot\cdot\cdot m_\nu \cdot\cdot\cdot m_N$ . For a small
number of modes and a reasonable number of phonons for each mode,
it can be easily diagonalized numerically. Given the eigenstates
of the Hamiltonian matrix (denoted by  $\left| k \right\rangle $),
 we can write down the electron Green's function\cite{Lifshitz}. In the
canonical ensemble,
\begin{equation}
G_{ij}(E)  = \frac{1}{Z} \sum\limits_{kk'}  (e^{- \beta E_k }+e^{-
\beta E_{k'}}) \frac{\left\langle k \right|a_i^+  \left| {k' }
\right\rangle \left\langle {k'} \right|a_j \left| k \right\rangle
} {E  - (E_k  - E_{k'} ) - i\eta } \label{Green's1}
\end{equation}
where $\beta=1/(k_B T)$ and $Z=\mbox {Tr } {\exp(-\beta H)}$. Although $\eta $
in Eq.(\ref{Green's1}) should be infinitesimal, we suppose it to be a small quantity
just for computational
purposes. Equation (\ref{Green's1}) can be rewritten as
\begin{equation}
G_{ij}(E) = \frac{1}{Z} \sum\limits_{kk'}  e^{ - \beta E_k } \left\{
\frac{\left\langle k \right|a_i^+  \left| {k' } \right\rangle
\left\langle {k'} \right|a_j \left| k \right\rangle } {E  - (E_k
- E_{k'} ) - i\eta } + \frac{\left\langle k \right|a_j  \left| {k'
} \right\rangle \left\langle {k'} \right|a_i^+ \left| k
\right\rangle } {E + (E_k  - E_{k' } ) - i\eta } \right\}.
\label{Green's2}
\end{equation}
In the first term inside the brackets, the state $|k\rangle$  has
one electron and the intermediate state $|k'\rangle$ has only
phonons. The opposite occurs in the second term. Admitting that it
costs an infinite energy to create a two-electron state, $\left|
{11\left\{ {m_\nu  } \right\}} \right\rangle$ cannot occur as the
intermediate state in Eq.(\ref{Green's2}). Therefore, we can write
for the diagonal elements of the Green's function:
\begin{equation}
G_{ii}(E)  = \frac{1}{Z} \sum\limits_{kk'} \left( e^{ - \beta E_k }+
e^{-\beta\sum_{\nu} m_{\nu}\hbar\omega_{\nu}} \right)
\frac{\left|C_i^k \left( {\{ m_\nu  \} } \right)\right| ^2} {E -
(E_k  -\sum\limits_{\nu  = 1}^N {m_\nu \hbar \omega _\nu  } ) -
i\eta } \label{Green's3}
\end{equation}
where $C_i^k \left( {\{ m_\nu  \} } \right)$  are the eigenvectors
expressed in terms of the basis vectors. From Eq.(\ref{Green's3}), the
fermion spectral density of states (SDS) is immediately obtained,
$$
\rho(E)=\frac {1}{\pi} \mbox{ Im }\sum_{i=1,2}{G_{ii}(E)}\equiv \sum_{i=1,2}{\rho _i},
$$
where $\rho _i$ is the partial SDS corresponding to the $i$-th bare electronic level.

The maximum number of phonons necessary to reproduce correctly several
lowest-energy eigenvalues (which are of importance for relevant
temperatures) depends on the coupling strengths but is not large.
Taking just 10 as the maximum number of phonons allowed in the
system, we obtained the eigenvalues coinciding with the analytical
results of  Ref.\cite{Mahan} (for $\alpha _{12}=0$ , $\alpha _{11}=
\alpha _{22}=0.2-0.3$) and formula (\ref{branches}) (for
$\alpha _{12} =0.2-0.3$ , $\alpha _{11}= \alpha _{22}=0$ and using
the RW approximation) to within $10^{-3}$meV. As an example, the
SDS spectra calculated for the latter case (but beyond the RW
approximation), are shown in
Fig.1.

\section{Emission and absorption spectra}
One can calculate the optical absorption and emission spectra by
using the Kubo formula for the frequency-dependent
 dielectric function\cite{Mahan}.
The imaginary part of the dielectric function, which describes the
absorption and emission properties, is related to the real part of
the frequency--dependent conductivity by the relation
\begin{equation}
\mbox {Im }\epsilon (\omega ) =\frac{4\pi}{\omega } \mbox {Re }\sigma (\omega ) .
\label{ImE}
\end{equation}
According to the Kubo formula,
\begin{equation}
\mbox {Re }\sigma (\omega ) =\frac{e^2}{m_0
V}\frac{1}{2\omega}\int\limits_{-\infty}^{\infty}{dt e^{i\omega t}
\langle j^+ (t) j(0) \rangle }.
\label{Kubo}
\end{equation}
We are interested in describing the transitions from an electron state (labelled
$0$) in the valence band (which we assume is not coupled to
phonons) to the states ($i=1,2$) in the conduction band, and {\it
vice-versa}. Thus, the current operator is defined as
\begin{equation}
j=\sum\limits_{i=1}^2 { p_{0i}\left( a_i^+a_0^{ } -a_0^+a_i^{ }  \right) }.
\end{equation}
The current-current correlation function is
\begin{equation}
\langle j^+ (t) j(0) \rangle=
\sum\limits_{i,i^{\prime}=1,2}{p_{0i}^* p_{0{i^{\prime}}} \left
\langle \left[ a_0^+ (t) a_i^{ }  (t) - a_i^+ (t) a_0^{ }  (t) \right]
\left[ a_0^+ a_{i^{\prime}}^{ }  - a_{i^{\prime}}^+ a_0 ^{ }  \right] \right
\rangle } .
\label{correlation}
\end{equation}
This expression can be developed into four expectation values.
However, there are only two terms which contribute to the sum in Eq.(\ref{correlation}).
This occurs because $a_i^{ }|k\rangle$ vanishes
if $|k\rangle$, an eigenstate of the Hamiltonian (\ref{Ham1}), is a state with zero electrons.
The non-zero terms are
$$
\langle a_0^+(t) a_i^{ }(t) a_{i^{\prime}}^+ a_0^{ } \rangle=
\frac{1}{Z}\sum\limits_{k}  e^{-\beta E_k} e^{iE_k t} \left \langle
k\left| a_0^+ a_i^{ } e^{-iHt}a_{i^{\prime}}^+ a_0^{ } \right| k \right\rangle
$$
and
$$
\langle a_i^+(t) a_0^{ }(t) a_0^+ a_{i^{\prime}}^{ } \rangle=
\frac{1}{Z}\sum\limits_{k}  e^{-\beta E_k} e^{iE_k t} \left \langle
k\left| a_i^+ a_0^{ } e^{-iHt}a_0^+ a_{i^{\prime}}^{ } \right| k \right\rangle
$$
where $a_i\equiv a_i(0)$.
These terms can be evaluated using the identity $1 =
\sum\limits_{k'}| k' \rangle \langle k' | $ by inserting the unity
between the $a$ operators. The result is:
$$
\langle j^+ (t) j(0) \rangle= \frac{1}{Z}\sum\limits_{k,
\{m_{\nu}\}} \sum\limits_{i,{i^{\prime}}} p_{0i}^*
p_{0{i^{\prime}}} e^{-\beta E_k} e^{i(E_k - E_0)t}
$$
\begin{equation}
\times \Bigl [ {C_{i^{\prime}}^k}^* (\{m_{\nu}\}) C_i^k
(\{m_{\nu}\})-{C_i^k}^* (\{m_{\nu}\}) C_{i^{\prime}}^k
(\{m_{\nu}\})\Bigl ].
 \label{correlation2}
\end{equation}
Using (\ref{correlation2}), performing the Fourier
transformation in (\ref{Kubo}) and substituting in Eq.(\ref{ImE}),
 we get the following expression
for the imaginary part of the dielectric function:
$$
\mbox {Im }\epsilon (\omega )=\left (\frac{2\pi e}{m_0\omega}\right )^2\frac
1{VZ^\prime} \sum\limits_{i,{i^{\prime}}=1,2} \Bigl \{p_{0i}^*
p_{0{i^{\prime}}}^{ }\sum\limits_{k, \{m_{\nu}\}} e^{-\beta \sum
_\nu m_{\nu}\hbar \omega_{\nu}}{C_i^k}^* (\{m_{\nu}\})
C_{i^{\prime}}^k (\{m_{\nu}\})
$$
\begin{equation}
\times \Bigl [\delta \Bigl (\omega -(E_k-\sum _\nu m_{\nu}\hbar
\omega_{\nu})\Bigl )+e^{-\beta (E_k-\sum _\nu m_{\nu}\hbar
\omega_{\nu})} \delta \Bigl (\omega +(E_k-\sum _\nu m_{\nu}\hbar
\omega_{\nu})\Bigl )\Bigl ]\Bigl \} \label{ImE2}
\end{equation}
where $Z^\prime= \sum\limits_{\{m_{\nu}\}} \exp {\Bigl [-\beta
\sum _\nu m_{\nu}\hbar \omega_{\nu}\Bigl ]}$. The terms in the
second line of Eq.(\ref{ImE2}) correspond to the absorption and
emission, respectively, of a photon of frequency $\omega$. Some
absorption and emission spectra calculated for a hypothetical QD are
presented in Figs.2-5 (the parameters are indicated on the figures),
 that demonstrate the effects of the
diagonal and off-diagonal coupling strength, inter-level spacing and
temperature on the optical properties of the dot. In these examples, we considered the
lower exciton level optically active and the upper one inactive ($p_2=0$).
 Such a situation occurs in spherical II-VI (e.g. CdSe) QDs ($1s_e1S_{3/2}$ and
 $1s_e1P_{3/2}$ states, respectively)\cite{Efros}.

\section{Interaction with acoustic phonons}
Interaction with confined longitudinal acoustic phonons, although mediated by a
different (namely, deformation potential instead of Fr\"ohlich)
mechanism, can be considered in the same way as that with optical
phonons and should result in series of closely spaced by still
isolated spectral peaks. However, for any kind of QDs embedded in a certain matrix,
 there must be a spectral region where acoustic phonons are essentially
delocalized and their energy varies in a continuous way. If the
velocities of sound for the QD and matrix material are not very
different, the density of acoustic phonon states should be similar
to that of bulk crystals and these states can be characterized by
a wavevector $\bf q$. This is the case of self-assembled QDs\cite{Cazayous2000} and
can be a reasonable approximation for a certain fraction of acoustic phonons in a spherical
 QD embedded in a dielectric matrix. In this section, we shall consider the interaction
of an electron localized in the dot with acoustic phonons which
are completely delocalized (see Appendix for details). The interaction, which occurs only
inside the QD, is weak for each phonon mode (since it contains a
factor $V_{QD}/V$, $V$ is the volume of the whole system). However, since the number of modes is virtually infinite,
 perturbation theory may fail. We shall avoid it. For this, we will have to
neglect coupling between different electronic levels through
acoustic phonons and consider a single electronic level coupled to
arbitrary number of phonon modes. This approximation corresponds to the independent boson model
\cite{Mahan}, which is normally considered for optical phonons (see Sec. III). Recently\cite{Krummheuer,Zimmermann2},
 it was applied to acoustic phonons in a QD. Based on the exact solution of this model, we shall propose a
 different procedure which will lead us directly to the self-energy function of the electron or exciton. Later,
 this self-energy will be re-interpreted for the optical-phonon polaron.

The Hamiltonian of a system consisting of one electronic level and a continuum of acoustic phonon modes
is:
\begin{equation}
H_2 = a^+ a \Bigl (\varepsilon _0 + \sum \limits_{\bf q} {g
_{\bf q} (b_{\bf q} ^ +  + b_{\bf q}^{})} \Bigl ) + \sum_{\bf q}
\hbar \omega _{\bf q } b_{\bf q} ^ +  b_{\bf q}^{}.
 \label{Ham2}
\end{equation}
It can be diagonalized by the transformation to new bosonic
operators\cite{Mahan}
\begin{equation}
B_{\bf q}^{} = b_{\bf q}^{}+\alpha _{\bf q } a^+ a \label{Trans}
\end{equation}
where $\alpha _{\bf q }=({g _{\bf q} }\slash { \hbar \omega _{\bf q
}})$. The energy spectrum is given by
\begin{equation}
E(\{N_{\bf q}\}) = \varepsilon _0 ^\prime + \sum_{\bf q} \hbar
\omega _{\bf q }N_{\bf q} \label{Spectr3}
\end{equation}
with $\varepsilon _0 ^\prime =\varepsilon _0^{ } -\sum_{\bf q} \alpha _{\bf
q}^2\hbar \omega _{\bf q }$. Eigenstates of the Hamiltonian
(\ref{Ham2}) can be expressed in terms of the pure phonon states,
\begin{equation}
\vert N_{\bf q}\rangle ^\prime=\sum_{m_{\bf q}} C_N^m
\vert m_{\bf q }\rangle.
\label{States}
\end{equation}
Acting with the operator $B_{\bf q}^{+}B_{\bf q}^{}$ on the
wavefunction (\ref{States}) we can find recurrent relations for
the coefficients $C_N^m$:
\begin{equation}
N=(m+\alpha _{\bf q}^{2}) (C_N^m)^2+\alpha _{\bf q }\Bigl [\sqrt
{m+1}C_N^{m+1}C_N^m+\sqrt {m}C_N^{m-1}C_N^m\Bigl ].
 \label{Recur}
\end{equation}
In linear (in the interaction with a single phonon mode $\bf q$) approximation, we get from
(\ref{States}) and (\ref{Recur}):
\begin{equation}
\vert N_{\bf q}\rangle ^\prime =p_{N_{\bf q}}\Bigl [\sqrt {N_{\bf q}} \alpha _{\bf q
}\vert N_{\bf q }-1\rangle+\vert N_{\bf q }\rangle +\sqrt {N_{\bf
q}+1} \alpha _{\bf q }\vert N_{\bf q }+1\rangle \Bigl ] \label{States1}
\end{equation}
where
$$p_{N_{\bf q}}=\frac 1{\sqrt {1+(2N_{\bf q}+1)\alpha _{\bf q}^{2}}}.$$
The one-electron Green's function corresponding to the Hamiltonian
(\ref{Ham2}) can now be calculated using Eq.(\ref{Green's1}),
$$
G(E)= \Bigl \langle \sum \limits_{\{m_{\bf q}\}} \frac {\prod \limits_{\bf
q}\{\delta _{m_{\bf q}, N_{\bf q }}+\alpha _{\bf q}^{2}[N_{\bf
q }\delta _{m_{\bf q}, N_{\bf q }-1}+(N_{\bf q
}+1)\delta _{m_{\bf q}, N_{\bf q }+1}]\}} {E  - \varepsilon
_0 ^\prime -\sum\limits_{\bf q} {\hbar \omega _{\bf q}(N_{\bf
q }-m_{\bf q})} - i\eta }
$$
\begin{equation}
\times \exp {\Bigl [-\sum \limits_{\bf q} {(2N_{\bf
q}+1)\alpha _{\bf q}^{2}}\Bigl ]}\Bigl \rangle
 \label{Green's4}
\end{equation}
where the angular brackets stand for thermodynamical average and the exponential factor appeared from
 the product $\prod\limits_{\bf q}p_{N_{\bf q}}$.
Taking into account only one-phonon processes, the
thermodynamical average approximately replaces $N_{\bf q }$-s with the
corresponding Bose factors $\bar N_{\bf q }=\Bigl (\exp {(\beta
\hbar \omega _{\bf q })}-1\Bigl )^{-1}$ and the Green's
function can be written as:
$$
G(E)= \Bigl \{ \frac 1 {E  - \varepsilon _0 ^\prime -i\eta } +\sum
\limits_{\bf q}\Bigl [ \frac {\alpha _{\bf q}^{2}\bar N_{\bf q }}{E  -
\varepsilon _0 ^\prime -\hbar \omega _{\bf q} - i\eta } +\frac
{\alpha _{\bf q}^{2}(\bar N_{\bf q}+1)}{E  - \varepsilon _0 ^\prime
+\hbar \omega _{\bf q} - i\eta } \Bigl ]\Bigl \}
$$
\begin{equation}
\times {\Bigl [1+\sum \limits_{\bf q} {(2\bar N_{\bf q}+1)\alpha _{\bf
q}^{2}}\Bigl ]}^{-1}.
 \label{Green's5}
\end{equation}
However, this equation (\ref{Green's5}) can be used only for quite a
weak interaction or at very low temperatures. Although the
interaction is weak for each phonon mode (all $\alpha _{\bf q }$ are
small), the number of modes is large and the effective
electron-phonon interaction is strong. A typical number of phonons
interacting with the electron can be estimated as $Q=(2\bar N_{\bf
q}+1)\alpha _{\bf q}^{2}$ and, for the model presented in Appendix,
$Q>>1$ for temperatures usually used in the experiments. In these
conditions, it is necessary to use the general equation
(\ref{Green's4}).

The evaluation of the Green's function from
Eq.(\ref{Green's4}) can be made using a Monte-Carlo procedure.
Instead of summing over all $3^{\mbox N}$ configurations $\{m_{\bf
q}\}$ (N is the number of acoustic phonon modes), one can generate some
$N_{MC}$ most probable configurations with $m_{\bf q}= (N_{\bf q}-1)$,
 $N_{\bf q}$, $(N_{\bf q}+1)$, occurring with the
probabilities $\alpha _{\bf q}^{2}\bar N_{\bf q}p_{\bar N_{\bf q}}^2$,
$p_{\bar N_{\bf q}}^2$, and $\alpha _{\bf q}^{2}(\bar N_{\bf
q}+1)p_{\bar N_{\bf q}}^2$, respectively. Averaging over a
sufficiently large $N_{MC}$ of such configurations, $G(E)$ is
obtained including all the many-phonon processes.

Let us define the electron self-energy taking into account the interaction with
acoustic phonons,
\begin{equation}
\Sigma (E) =E-G^{-1}(E). \label{SE1}
\end{equation}
In the one-phonon approximation, the explicit expression for the
self-energy is:
$$
\Sigma (E)= \varepsilon _0 ^\prime -(E-\varepsilon _0 ^\prime )
\sum \limits_{\bf q}{\alpha _{\bf q }^2(2\bar N_{\bf q}+1)}
$$
\begin{equation}
+(E-\varepsilon _0 ^\prime )^2 \sum \limits_{\bf q} \Bigl [\frac
{\alpha _{\bf q}^{2}\bar N_{\bf q }}{E  - \varepsilon _0 ^\prime -\hbar
\omega _{\bf q} - i\eta } +\frac {\alpha _{\bf q}^{2}(\bar N_{\bf
q}+1)}{E  - \varepsilon _0 ^\prime +\hbar \omega _{\bf q} - i\eta
} \Bigl ].
 \label{SE2}
\end{equation}
The spectral dependence of the self-energy corresponding
to the Green's function (Eq.(\ref{Green's5})) obtained using the Monte Carlo
procedure is presented in Figs.6,7. For comparison, we also calculated the self-energy
using Eq.(\ref{SE2}). The interaction constants $\alpha _{\bf q }$
derived in the Appendix were used in these calculations and the
material parameters were taken as for CdSe, except for the deformation potential constant $a_c$,
which was taken about 2 times smaller than the bulk value of the relative volume
 deformation potential between the valence and conduction bands \cite{Yu-Cardona}. Such
a choice is justified by the fact that only a fraction of acoustic phonons can be described by
 a propagating wave assumed in the Appendix.

Assuming that the broadening of the electronic levels produced by
the interaction with acoustic phonons is small compared to the
electronic level spacing, it is possible to include this
interaction in our scheme of consideration of polaronic states
proposed in the previous sections.
Let us consider now the full Hamiltonian, which includes the
interactions with both optical and acoustic phonons,
$$
H_3 = \sum\limits_{i = 1}^2 {\Bigl ( \varepsilon _i + \sum
\limits_{\bf q} {g _{\bf q}^{(i)} (b_{\bf q} ^ +  + b_{\bf
q}^{})} \Bigl ) a_i^ +  a_i^{ } } + \sum \limits_{ij}^2 {\sum
\limits_{\nu  = 1}^N {g _{ij}^{\scriptstyle  \hfill \atop
\scriptstyle \nu  \hfill} } a_i^ +  a_j^{ } \left( {b_\nu ^ +
 + b_\nu ^{ } } \right)}
$$
\begin{equation}
+ \sum_{\bf q} \hbar \omega _{\bf q } b_{\bf q} ^ +  b_{\bf q}^{}
+ \sum \limits_{\nu  = 1}^N {\hbar \omega _\nu  } {b_\nu ^ + b_\nu
^{ }  } \label{Ham4}
\end{equation}
where $\alpha _{\bf q}^{(i)}$ denotes the acoustic phonon coupling
constant for electronic level $i$. The Hamiltonian (\ref{Ham4})
can be rewritten in terms of the polaron states $|k \rangle$ by
introducing the corresponding (fermionic) annihilation
and creation operators $A_k^ + , A_k^{ }   $. The electron-LO-phonon Hamiltonian
(\ref{Ham1}) is then reduced to
$$
H_1 = \sum _kE_k A_k^ + A_k^{ }
$$
Using the
expansion of the bare electronic states in terms of the polaron ones,
$$
\left| i \right \rangle =\sum _k \sum _{\{m_\nu\}}\Bigl (C_i^k \left( {\{
m_\nu  \} } \right)\Bigl )^* \left| k \right \rangle ,
$$
the term in (\ref{Ham4}) representing the interaction with
acoustic phonons can be written as
$$
\sum \limits_{i=1,2} \sum \limits_{\bf q} {g _{\bf q}^{(i)}
(b_{\bf q} ^ +  + b_{\bf q}^{})}\sum _{k,k^\prime} \sum
_{\{m_\nu\}} \Bigl (C_i^k \left( {\{ m_\nu  \} } \right)\Bigl
)^*C_i^{k^\prime } \left( {\{ m_\nu  \} } \right) A_k^ +
A_{k^\prime }^{ }
$$
\begin{equation}
=\sum \limits_{\bf q} {g _{\bf q}^{} (b_{\bf q} ^ +  + b_{\bf
q}^{})}\sum _{k} A_k^ +  A_{k}^{ } +\sum \limits_{\bf q} {\Delta
g _{\bf q}^{} (b_{\bf q} ^ +  + b_{\bf q}^{})} \sum
_{k,k^\prime } t_{k,k^\prime }A_k^ +  A_{k^\prime }^{ }
\label{Term}
\end{equation}
where $g _{\bf q}^{} =(g _{\bf q}^{(1)}+g _{\bf
q}^{(2)})/2$,
 $\Delta g _{\bf q}^{} =(g _{\bf q}^{(2)}-g _{\bf q}^{(1)})$, and
$$
t_{k,k^\prime }=\sum _{\{m_\nu\}} \Bigl \{ \left (C_1^k \left( {\{
m_\nu  \} } \right)\right )^*C_1^{k^\prime } \left( {\{ m_\nu  \}
} \right)- \left (C_2^k \left( {\{ m_\nu  \} } \right)\right
)^*C_2^{k^\prime } \left( {\{ m_\nu  \} } \right) \Bigl \}.
$$
Neglecting the last term in (\ref{Term}), that is, assuming that
the coupling to acoustic phonons is approximately the same for both electronic levels,
 we come to the Hamiltonian
\begin{equation}
H_3 =\sum _{k} A_k^+ A_k^{ } \Bigl (E _k + \sum
\limits_{\bf q} {g _{\bf q} (b_{\bf q} ^ +  + b_{\bf q}^{})}
\Bigl ) + \sum_{\bf q} \hbar \omega _{\bf q } b_{\bf q} ^ + b_{\bf
q}^{} .
\label{Ham5}
\end{equation}
It can be diagonalized exactly in the same way as (\ref{Ham2}). The poles of the electron
 Green's function (\ref{Green's3}) (as well as the absorption and emission peaks) are shifted according to the replacement
$E_k\rightarrow E_k^{\prime}$,
$$
E_k^{\prime}=E_k -\sum_{\bf q} \alpha _{\bf
q}^2\hbar \omega _{\bf q }+
\Bigl \langle \sum_{\bf q}
\hbar \omega _{\bf q }(N_{\bf q }-m_{\bf q})\Bigl \rangle ,
$$
where, as before, $(N_{\bf q }-m_{\bf q})$ can be considered a random variable taking the values 1, 0 and -1
with the probabilities $\alpha _{\bf q}^{2}\bar N_{\bf q}p_{\bar N_{\bf q}}^2$,
$p_{\bar N_{\bf q}}^2$, and $\alpha _{\bf q}^{2}(\bar N_{\bf
q}+1)p_{\bar N_{\bf q}}^2$, respectively.
 The result is a similar broadening and a downward shift of all the polaron states
 contributing to the spectral density of states, absorption and emission, just like it happens in the one-level
independent boson model\cite{SR}.
It is equivalent to ascribing a
spectral variable--dependent self-energy, $\Sigma (E-E_p)$ (given by Eqs.(\ref{SE1}) or (\ref{SE2})),
 to each pole of (\ref{Green's3}). As before, we can re-interpret this result for exciton--polaron, under the condition
 that we have no more than one exciton per QD.
An example of spectra showing this effect is presented in Fig.8.
The non-diagonal term with $t_{k,k^\prime }$ leads to acoustic
phonon--mediated mixing of the polaron states. It will be
considered in a future work.

\section{Discussion}
Let us start the discussion by emphasizing that only in a hypothetical case when the phonon--mediated coupling
 of the lowest energy exciton state to
 all other states can be neglected (for example, in the limit of extremely strong confinement, such that
 $\Delta >> \hbar \omega _0$, or if $\alpha _{12}$ is small because of the symmetry of the corresponding wavefunctions),
 one can expect to observe Franck-Condon progressions in the emission and absorption
spectra associated with this state\cite{SR}. This is the only case when the Huang-Rhys parameter describes the
 spectra in entirety. As it is obvious from Eq.(\ref{alpha}), the {\it diagonal} coupling is proportional to the
 (integrated) difference between the electron and hole charge densities, which is non-zero mostly due to the fact
 that the conduction and valence bands in II-VI and III-V materials have different symmetry.
 Even if some further effects (like the presence of defects or strain, for example) contribute to the separation of the electron
 and hole clouds in space\cite{Heitz99}, one can hardly expect an order of magnitude increase of this parameter, compared to the
 calculated values cited in the Introduction. It means that the intensity of the LO phonon sidebands must be rather small
 and monotonically decreasing with the increase of the number of phonons.

However, in most cases the phonon--mediated interlevel coupling, at least between the subsequent exciton states, should be important.
The off-diagonal interaction constant is likely to exceed the diagonal ones. For instance, if the two exciton levels are
different by the hole state (e.g., $1s_e1S_{3/2}$ and $1s_e1P_{3/2}$ states in a spherical II-VI QD), it is easy to see from
Eq.(\ref{alpha}) that (neglecting the Coulomb interaction between the electron and hole),
\begin{equation}
g _{12}^{\nu}  = e\int {\Psi _{1h}^{*} (\vec r_h
)\phi _{\nu} (\vec r_h )\Psi _{2h} (\vec r_h )d\vec r_h},
\label{alpha2}
\end{equation}
that is, there is no compensation effect for this interaction. The spectra calculated assuming only the off-diagonal interaction
(Figs.1,2) show characteristic features known as Rabi splitting, which is obtained already
 in the RW approximation (see Eq.(\ref{branches}))
 and has been observed experimentally\cite{Hameau}. Contrary to the RW approximation, the exact results also show
 a downwards shift of the spectral lines (see Fig.1. In the RW approximation, the most intense peak should be
 situated exactly at $E-E_0=0$). Even if the upper level is dipole forbidden, its presence manifests itself by the
 spectral features seen in Fig.2. Note that the off-diagonal coupling is not resonant as one might expect thinking
 in terms of the Fermi's Golden Rule. The manifold produced by the coupling persists even for quite a large detuning
$\Delta^\prime >\hbar \omega _0$. Curiously, it is reminiscent of sidebands which might appear due to interaction
 with confined acoustic phonons.

When both diagonal and off-diagonal interactions are present, the
effect is not just a sum of those owing to each of them, but
also some additional spectral features appear (see Figs.3-5 showing the effect of different parameters).
The stronger are the interactions, the richer the structure of the spectra. It should also be noted that the symmetry
between the absorption and emission spectra, which is characteristic of the one-level
independent boson model (see the upper panel of Fig.5), disappears when the off-diagonal
 interaction is included. This holds also at low temperatures where a small number of spectral features are seen (Fig.3, lower panel).
 Those below $E_0$ in the emission spectrum are due to the diagonal coupling to the lowest energy exciton state.
 The one seen in the absorption originates from the inter-level coupling.

 Let us turn to the acoustic phonons. The explicit calculation leads to a self--energy which depends on the spectral
 variable in a sophisticated way. Accordingly, the line shape is quite different from the simple
 Lorentzian\cite{Krummheuer,Zimmermann2}.
 The noise seen in the calculated spectra of the self--energy (Figs.6,7) is due to the Monte Carlo procedure used when
each configuration ${\{m_{\bf q}\}}$ of phonons produces a $\delta $-function and the number of Monte Carlo runs is finite
($N_{MC}=20000$ for Figs.6,7). The imaginary part of the self--energy vanishes at $\varepsilon _0 ^\prime$, so, the zero-phonon
 line (ZPL) at low temperatures remains a sharp $\delta$-like  resonance situated between two asymmetric phonon sidebands,
 in accordance with experiment\cite{Palinginis,Warming} and previous theoretical studies\cite{Krummheuer,Zimmermann2}.
Only for high temperatures the SDS generated by a single electronic level takes a bell-like shape similar to the Lorentzian. The calculated results also
confirm the statement concerning the importance of the multiphonon processes which increase in smaller QDs (Fig.6).
Turning to the exciton--polaron spectrum (Fig.8), the acoustic phonon--related broadening smears the discrete structure associated with optical phonons
The multiplicity of the confined optical phonon modes also helps this. Below, we shall explain some previously published
 experimental findings, which were not clearly understood before, in terms of our calculated results.

{\bf ``Strange" phonon replica}

 An experimental evidence of spectral features whose distance from ZPL is smaller than the LO phonon energy was found
  in several works\cite{Bissiri,Ignatiev,Fafard,Odnolyub}. Thinking in terms of the Franck-Condon progression, such peaks were
  attributed to additional (interface or disorder-activated acoustic) phonon modes, for some reason strongly coupled to the
  exciton\cite{Odnolyub}. As it can be seen from Figs.3-5, in the presence of inter-level coupling, there are many
  spectral features that are not separated from ZPL (the most intense peak in all spectra) by (a multiple of) the optical
  phonon energy.
  Some of them persist at low temperatures. Such ``strange" replica can be generated by a {\it single} phonon mode
  (long-wavelength LO phonon, neglecting its confinement) and their spectral positions are determined by the e-ph coupling constants.
Thus, there is no need to invoke extra modes (for which it would be hard to justify their strong coupling to the exciton) in order to
 explain the discussed spectral features. The phonon--mediated inter-level coupling provides a simpler and more plausible explanation.

{\bf Anomalously strong 2LO phonon satellite}

As mentioned in the Introduction, several groups observed
 apparently phonon--related features in the absorption (but not in the emission) spectra of resonantly excited self--assembled
QDs\cite{Fafard,Schmidt,Lemaitre,Toda,Heitz}. Given the relative weakness of the exciton-phonon interaction in such dots ($S<<1$),
the most difficult to explain was the fact that the most intense satellite was not the 1LO phonon sideband
 but rather the 2LO or 3LO one. This effect was qualitatively understood as being due to a resonant coupling of the
 corresponding phonon replica of the ground state to the excited state of the exciton\cite{Lemaitre}. Considering a QD ensemble, one can think
 in terms of a multiphonon ``filtering" of inhomogeneously distributed excited states\cite{Toda}. However, when a single QD
 spectroscopy is used\cite{Lemaitre}, such a resonance of the inter-level spacing with a multiple of the phonon energy
 is rather improbable. (Even though there are several confined optical phonon modes with slightly different energies,
  capable of considerable coupling to the exciton, this dispersion is only of the order of 1meV). The assumption of a
  reasonably strong off-diagonal coupling of the lowest and higher exciton states (not in a close resonance with a certain
  number of phonons) can account for anomalously strong $n$LO phonon sidebands. Fig.9 shows how peaks separated by energies approximately equal to that of the optical phonon can appear
 in the absorption spectra. We took a value of $\Delta$=55meV, typical of InAs/GaAs self-assembled QDs, and considered two
  optical phonon modes with the energies of 32 and 30 meV in order to simulate the experimental result of Lemaitre {\it et al}\cite{Lemaitre}.
  (The two modes can be interpreted as LO phonon in pure InAs and InAs-like LO phonon in the InGaAs alloy near the QD boundary, respectively).
  The values of $\alpha ^{\nu}_{12}=0.2$ and 0.15 used in this calculation do not look extraordinarily high taking into account the argument presented
 above (see Eq.(\ref{alpha2})) and the possible contribution of the optical deformation potential mechanism in this interaction.
 As it can be seen from Fig.9, the second strong absorption peak is {\it not} separated from ZPL by the energy $\Delta$
 and can be called ``2LO satellite", since the weaker 1LO and 3LO ones also show up in the spectrum. The calculated spectrum is
  in a good agreement with the experimental observation of Ref.\cite{Lemaitre}.

{\bf Homogeneous broadening of absorption lines}

In structures of higher dimensionality, the temperature dependence of the homogeneous linewidth is described by the following
 relation,
 \begin{equation}
\Gamma (T) -\Gamma (0)=\gamma _{ac}n_{ac}(T)+\gamma _{LO}n_{LO}(T)
\label{gamma}
\end{equation}
where $\gamma _{ac}$ and $\gamma _{LO}$ are some constants and $n_{ac}$ and $n_{LO}$ the Bose factors for the
 acoustic and optical phonons, respectively.
 There is no reason for Eq.(\ref{gamma}) be a good approximation for QDs, because it relies on the existence of
  a continuum of electronic states and also is limited to one-phonon scattering processes. Still it was used in
  \cite{HR1,Rakovich} and some other publications to qualitatively describe the dependence $\Gamma (T))$ extracted
   from experimental absorption spectra of spherical QDs. It was found that this dependence,
  in the region from 77 to 300K, is much stronger than linear, similar to that predicted by the second term
  in Eq.(\ref{gamma}), suggesting that there is some broadening related to the optical phonons. As it has been
  pointed out in the Introduction, interaction of localized electrons with optical phonons produces only additional discrete
   spectral lines. Consequently, the optical phonon related part of the broadening should not be understood literally, strictly speaking, the broadening occurs only because of the
 continuum of acoustic phonons. Our calculated Fig.8, where the interaction constants were taken as for a typical
 spherical CdSe QD studied in  Ref.\cite{Rakovich}, demonstrates this effect. The interaction with several optical phonon
 modes results in a series of closely spaced discrete peaks (see inset in Fig.8)) which is camouflaged by the acoustic phonons.
 The effective homogeneous broadening of the spectral lines may apparently depend on the equilibrium number of
 acoustic and optical phonons. Nevertheless, Eq.(\ref{gamma}) is too simple to describe this effect even qualitatively,
  as found in Ref.\cite{Bayer} using single QD spectroscopy.

Owing to the apparent similarity with systems of higher dimensionality, some authors associate the broadening
 discussed above with the optical phonon--assisted carrier relaxation. However, as it has been pointed
out\cite{Inoshita,Levetas}, this is not a lifetime (or irreversible scattering) effect.
Indeed, the polaron states considered here are stationary
states. The phonon-assisted relaxation can probably occur through
an anharmonic decay of the phonons participating in the formation of the
polaronic states, as suggested in Refs.\cite{Levetas,Li} and  calculated in Ref.\cite{Verzelen}.
Our consideration of the polaron spectrum in the presence of acoustic phonons shows,
 however, another possibility, namely, the acoustic phonon--mediated transitions between
different polaron states  owing to the non-diagonal term in (\ref{Term}).
 The next step should consist of analyzing the efficiency of this mechanism. It is worth noting at this point that,
 contrary to the opinion of the authors of Ref.\cite{Jasak}, such acoustic phonon-mediated transitions in the polaron
  spectrum should not be subject to phonon bottleneck, because there are plenty of polaron states due to different optical phonon modes
 differently coupled to the exciton. Many of them (which not necessarily show up in the spectra corresponding to thermal equilibrium) are
 separated by small energies within the band of acoustic phonons efficiently interacting with the polaron.

{\bf Up-converted PL}

The up-converted or anti-Stokes photoluminescence (ASPL) is the emission of photons with energies higher than the excitation
 energy ($E_{exc}$). This effect was observed for colloidal II-VI QDs and discussed in several recent
  publications\cite{Poles,Rak,Wang}. Additional information can be found in Ref.\cite{Filan}. The ASPL occurs when an ensemble of
  QDs is excited at the very edge of the absorption spectrum, below the normal (i.e. excited with high energy photons) PL
  band. The principal experimental facts concerning this effect are the following:

  (i) The ASPL intensity increases linearly with the excitation power \cite{Rak}, which can be rather low.

  (ii) The blue (anti-Stokes) shift between the ASPL peak and $E_{exc}$ does not depend significantly upon the QD size if
  $E_{exc}$ is chosen proportionally to the absorption peak energy (which depends upon the size)\cite{Rak}.
   However, the shift increases with temperature and can range approximately from 20 to 150meV\cite{Filan}.

   (iii) If $E_{exc}$ increases (approaching the absorption peak) the ASPL also moves continuously towards higher energies\cite{Rak}.
   Its intensity increases, and finally the spectrum transforms into the normal PL band\cite{Filan}.

   (iv) The ASPL intensity increases strongly with temperature.

   Possible ASPL mechanisms were discussed in Refs.\cite{Poles,Rak,Wang}, but no definite conclusion was made. According to (i),
   processes like two-photon absorption and (since the possibility of emergence of more than one exciton per QD is negligible)
   Auger excitation can be excluded in this case. Therefore, it was suggested that incident photons excite electrons to
   some intermediate sub-gap states from which they eventually proceed to the higher  energy (luminescent) states through
   the thermal effect. Some obscure surface states (SS's) for which, as admitted by the authors of Ref.\cite{Wang}, there is no
   any direct evidence, were suggested as those responsible for the sub-gap absorption\cite{Rak,Wang}.

   From our point of view, the participation of sub-gap SS's (even if they exist in the passivated colloidal QDs) is
   unlikely in virtue of the experimental results (ii) and (iii). In fact, it would require the SS's energy have approximately
   the same dependence on the QD size as that of the confined electronic states in the dots, in obvious contradiction with what
   should be expected from the general point of view. At the same time, there exist naturally formed sub-gap states, which are
   separated from the fundamental absorption line (i.e., ZPL) by energies which are only weakly dependent on the QD size. These
   are the red-shifted optical phonon replica. Fig.10a presents the lower--energy side of a calculated QD absorption spectrum
 showing two sub-gap bands (designated as ``-1LO" and ``-2LO") through which the dot can be excited. The excited QD then will emit
 a photon most likely having the ZPL energy. The probability of such an up-conversion process increases with temperature because
 so does the integrated intensity of ``-1LO" and ``-2LO" absorption bands (shown calculated in Fig.10b). The experimental
 temperature dependence of the ASPL intensity measured in Ref.\cite{Rak} can be understood taking into account that, at
 a certain temperature, further red--shifted satellites (whose intensity depends more strongly on the temperature) become more
  efficient. The situation is complicated by the distribution of the QD size, so that $E_{exc}$ can match different
  ``$-n$LO" bands of dots of several different sizes. This agrees qualitatively with the experimentally observed increase in the
  anti-Stokes shift with temperature\cite{Filan}. Modeling of the ensemble effects involved in the ASPL phenomenon is beyond
  the scope of this paper and will be considered in a future work. Nevertheless, we believe that our model explains, at least
  qualitatively, the principal experimental facts concerning the ASPL.

In conclusion, we proposed a non-perturbative approach for the calculation of the
polaronic effects in QDs, which allows the consideration of electronic
levels coupled through the interaction with several confined
phonon modes. Using this approach, we were able to show that the
polaronic effects are significant even when the interlevel spacing
is quite away from resonance with the optical phonon energies. We demonstrated that this opens the possibility
 to account for some previously published and not clearly understood experimental data. We also presented an explicit
 calculation of the spectral line shapes because of the diagonal interaction of acoustic phonons
with the original electronic levels and suggest that the same interaction may be responsible for relaxation
 of the polaron entity to lower energies.

\section{Acknowledgments}
This work was supported by the FCT (Portugal) through project FCT-POCTI/FIS/10128/98, ICCTI-CNPq (Portugal-Brazil) Cooperation
Program and Brazilian agencies CNPq and FAPERJ. The authors are grateful to S.Filonovich for discussions of the ASPL
 and to T.Warming for providing
 the manuscript of Ref.\cite{Warming} prior to publication.

\newpage
\appendix*
\section{Electron-acoustic-phonon coupling rates}

The following simple model was used to estimate the rates $\alpha _{\bf
q}$ of electron coupling to longitudinal acoustic phonons. Assuming that LA
phonons are completely delocalized, that is,
 neglecting acoustic impedance at the interface between the QD and matrix, the displacement can be written as
$$
u_z=\Bigl (\frac  \hbar {2\omega _{q} \rho_0V } \Bigl )^{1/2}\mbox
{Re } e^{i(qz-\omega _{q}t)},
$$
where $\rho _0$ is the density. For the deformation potential
interaction mechanism, the dimensionless coupling constant is \cite{Yu-Cardona}
$$
\alpha _{\bf q}=\frac {a_c}{\hbar \omega _{\bf q}}\int |\psi |^2\mbox {div}
{\bf u}({\bf q})dV,
$$
where $a_c$ is the bulk deformation potential and $\psi $ the
electron wavefunction. Considering the lowest conduction band
state in a spherical QD of radius $R$, with infinite barriers, the
wavefunction is \cite{Efros}:
$$
\psi =\sqrt {\frac 1{2\pi R}}\frac {\sin{(\pi r/R)}}R \qquad (r\le
R).
$$
Using these expressions, we obtain
\begin{equation}
\alpha _{\bf q}=a_c\Bigl (\frac  1{2\hbar \omega _{q} \rho_0Vc_l^2}\Bigl
)^{1/2}I(qR)
\label{zq}
\end{equation}
where $c_l$ is the longitudinal sound velocity,
\begin{equation}
I(y)=2\pi^2\int _0^1j_0(y\cdot x)j_0^2(\pi x)x^2dx
\label{integr}
\end{equation}
and $j_0$ is the spherical Bessel function. The integral in (\ref{integr}) can be expressed in terms of the integral
 sinus as
$$
\frac 12\{\mbox {Si}(y)-\frac 12[\mbox {Si}(y+2\pi)+\mbox {Si}(|y-2\pi|)]\}.
$$
The function $I(y)$
decreases from 1 for $y=0$ to nearly zero for $y\ge 5$.

%\begin{references}

\newpage
\section*{Figures}
\begin{figure}[ht]
%Fig.1
\begin{center}
\includegraphics[height=12cm]{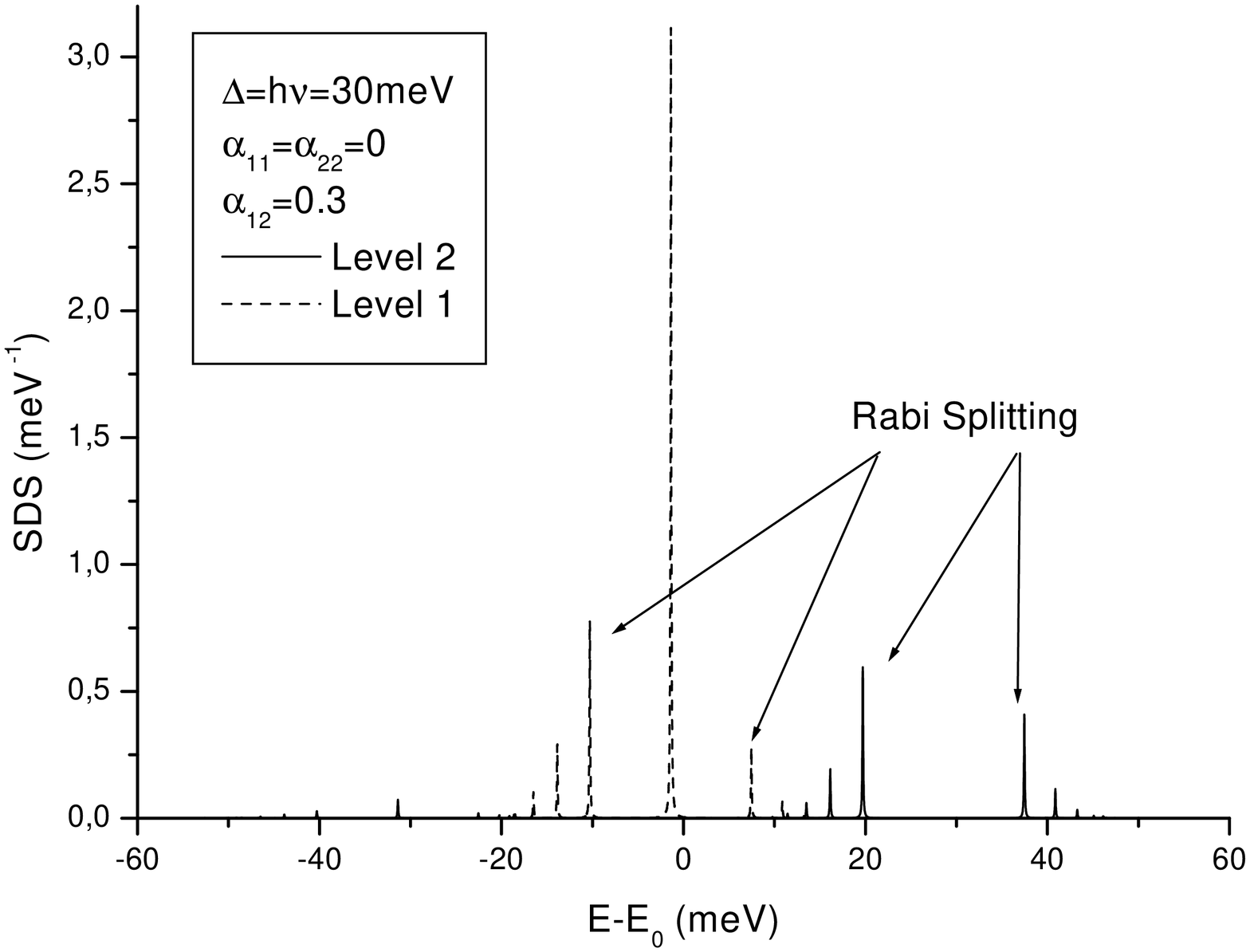}
\caption{ Calculated partial spectral densities of states
corresponding to two electronic levels separated by $\Delta=30meV$
interacting with one optical phonon mode with
$\hbar \omega _0=30meV$ (that is, $\Delta^\prime =0$).
Only non-diagonal coupling was included ($\alpha _{12}=0.3$).
Note the Rabi splitting in the polaron states and their shift to
the lower energies.}
\end{center}
\label{figure01}
\end{figure}

\begin{figure}[ht]
%Fig.2
\begin{center}
\includegraphics[height=12cm]{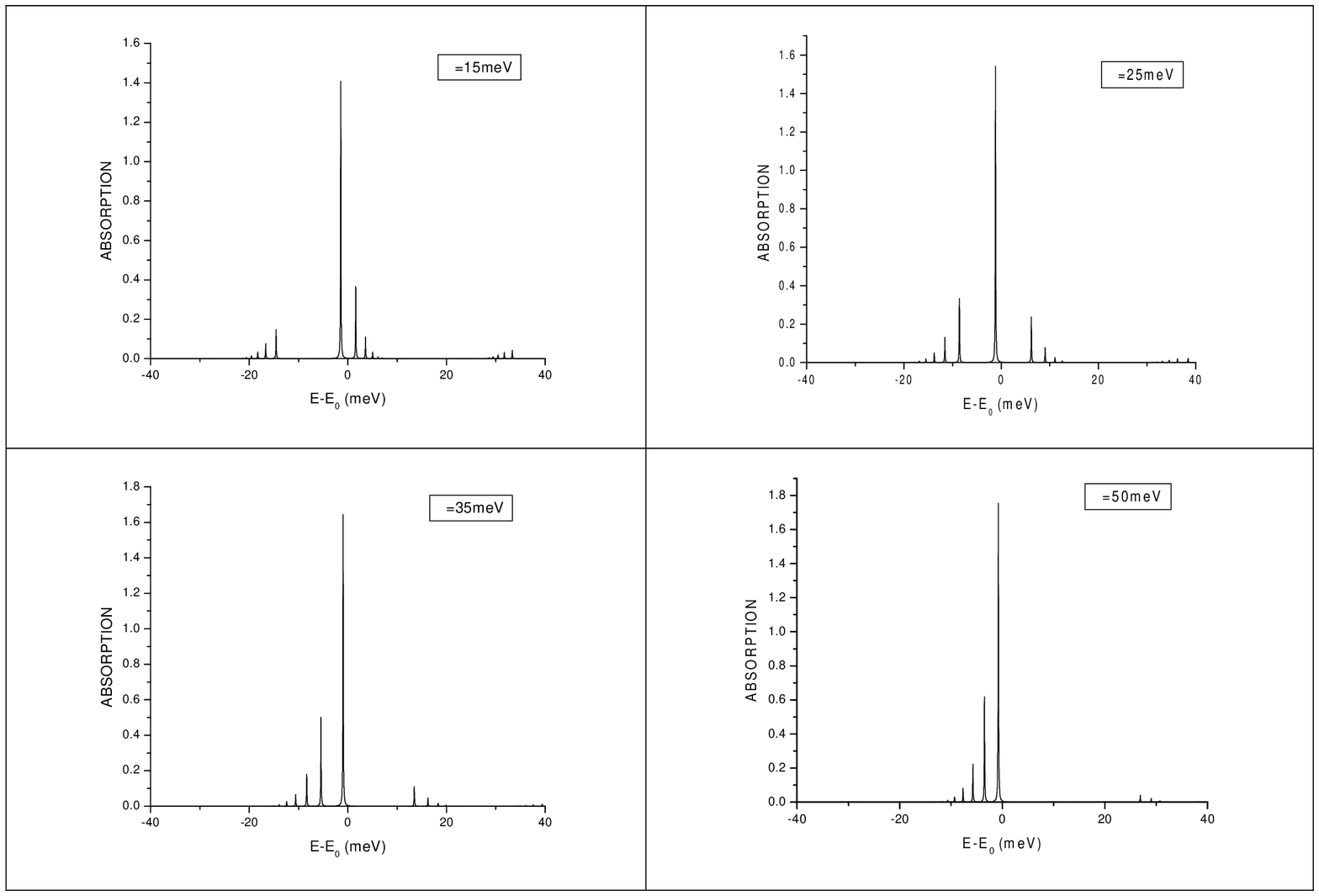}
\caption{ Calculated absorption spectra for different values of
level spacing as indicated. All the other parameters are the same
as for Fig.1. }
\end{center}
\label{figure02}
\end{figure}

\begin{figure}[ht]
%Fig.3
\begin{center}
\includegraphics[height=12cm]{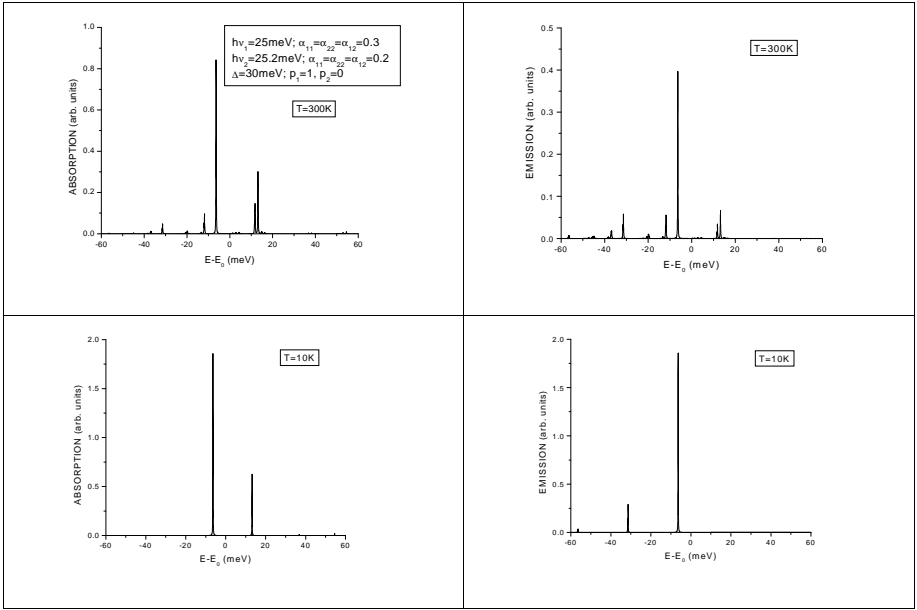}
\caption{ Absorption (left) and emission (right) spectra for a
system of two exciton levels separated by $\Delta =30meV$
calculated for different temperatures. The excitons interact with
two optical phonon modes of the energies $\hbar \omega _1=25meV$
and $\hbar \omega _2=25.2meV$. The electron-phonon coupling
constants are indicated on the figure. The origin of the energy
axis is chosen at the energy of the lowest exciton state ($E_0$).}
\end{center}
\label{figure03}
\end{figure}

\begin{figure}[ht]
%Fig.4
\begin{center}
\includegraphics[height=12cm]{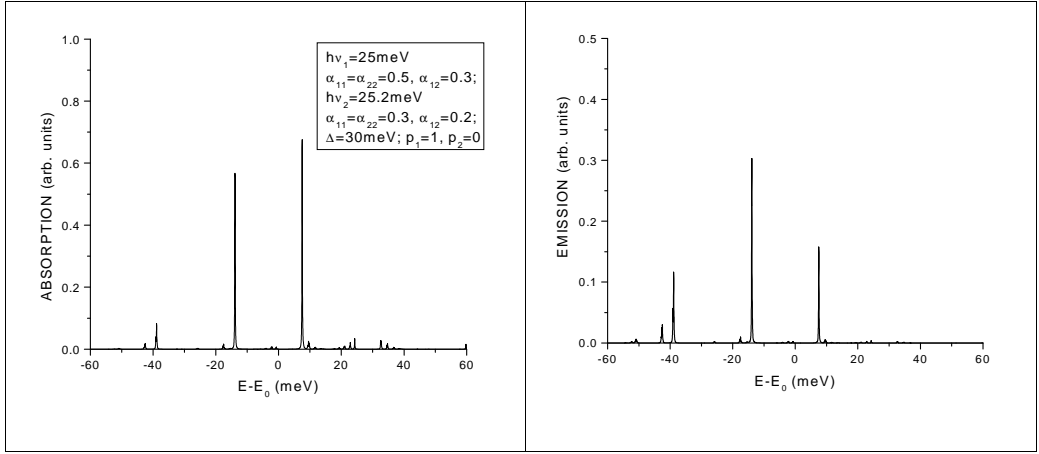}
\caption{ Absorption (left) and emission (right) spectra
calculated for different (compared to Fig.3) values of the
diagonal coupling constants. }
\end{center}
\label{figure04}
\end{figure}

\begin{figure}[ht]
%Fig.5
\begin{center}
\includegraphics[height=12cm]{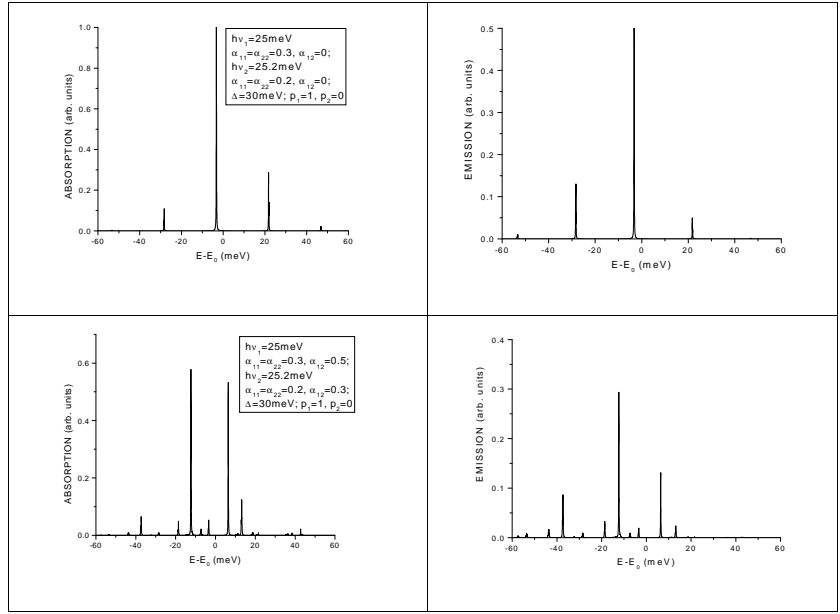}
\caption{ Absorption (left) and emission (right) spectra
calculated for different values of the non-diagonal coupling
constant (see the explanation to Fig.3). }
\end{center}
\label{figure05}
\end{figure}

\begin{figure}[ht]
%Fig.6
\begin{center}
\includegraphics[height=12cm]{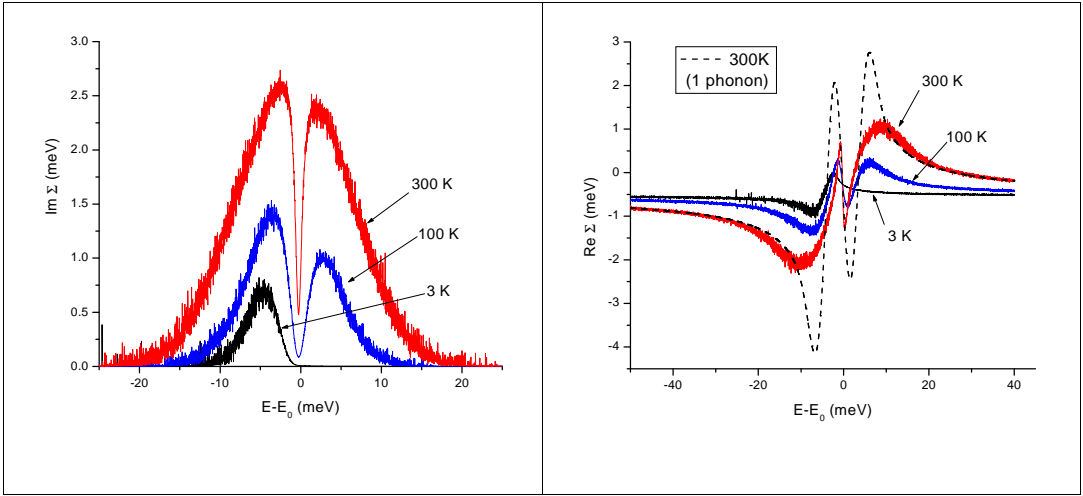}
\caption{ Imaginary (left) and real (right) parts of the
self--energy for a single electronic level interacting with
acoustic phonons, calculated for different temperatures using the
Monte-Carlo technique explained in the text. The dashed curve was
plotted using the one-electron approximation (Eq.(22)). The
coupling constants were calculated as explained in the Appendix.
The necessary material parameters were taken as follows:
$\rho _0=5.8g/cm^3$, $c_l=3.8km/s$ and
$a_c=3.5eV$. The QD radius $R$=2nm.}
\end{center}
\label{figure06}
\end{figure}

\begin{figure}[ht]
%Fig.7
\begin{center}
\includegraphics[height=12cm]{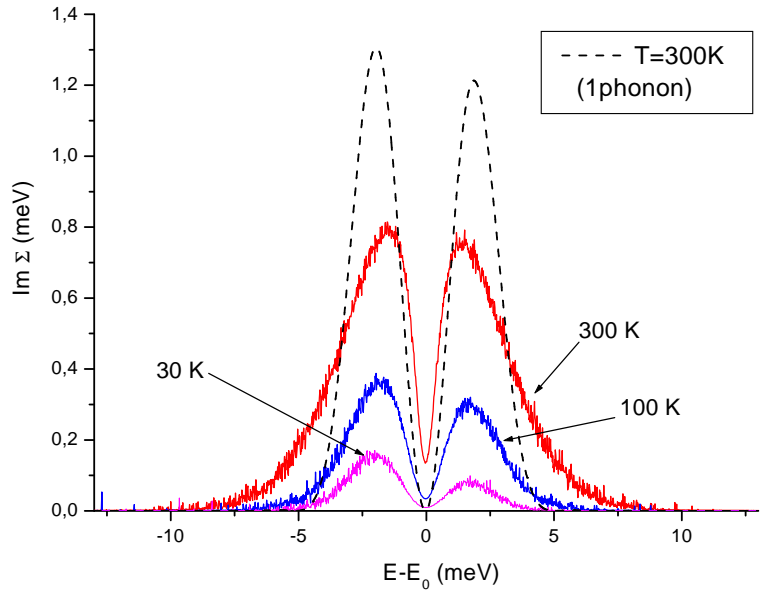}
\caption{ Imaginary part of the self--energy calculated for a
$R$=4nm QD as explained for Fig.6. The dashed curve was plotted
using one-electron approximation (Eq.(22)). }
\end{center}
\label{figure07}
\end{figure}

\begin{figure}[ht]
%Fig.8
\begin{center}
\includegraphics[height=12cm]{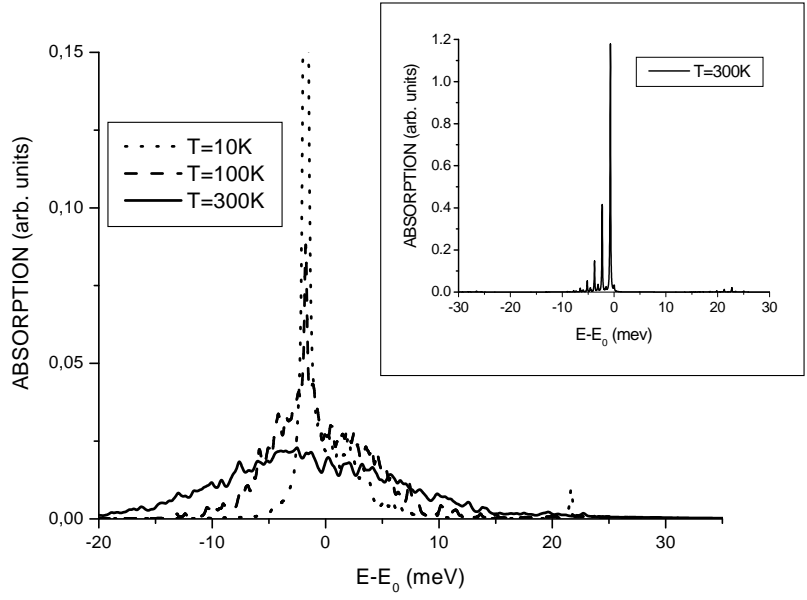}
\caption{ Absorption spectra calculated for an $R$=2nm CdSe QD
including both optical and acoustic phonons for three different
temperatures. Two optical phonon modes of the energies
$\hbar \omega _1=25meV$ and $\hbar \omega _2=25.8meV$ and
dimensionless interaction constants (divided by the corresponding
phonon energies)
$\alpha _{11}^{1}=-0.015$,
$\alpha _{22}^{1}=0.027$,
$\alpha _{12}^{1}=0.295$;
$\alpha _{11}^{2}=0.063$,
$\alpha _{22}^{2}=0.053$,
$\alpha _{12}^{2}=0.095$
calculated in Ref.\cite{Vasilevskiy2} were used.
The acoustic phonon parameters were taken as for Fig.6.
The level spacing of $\Delta =80meV$ was assumed.
The inset shows the $T=300$K spectrum without acoustic
phonon contribution (at $T=10$K only the main peak is seen).}
\end{center}
\label{figure08}
\end{figure}

\begin{figure}[ht]
%Fig.9
\begin{center}
\includegraphics[height=12cm]{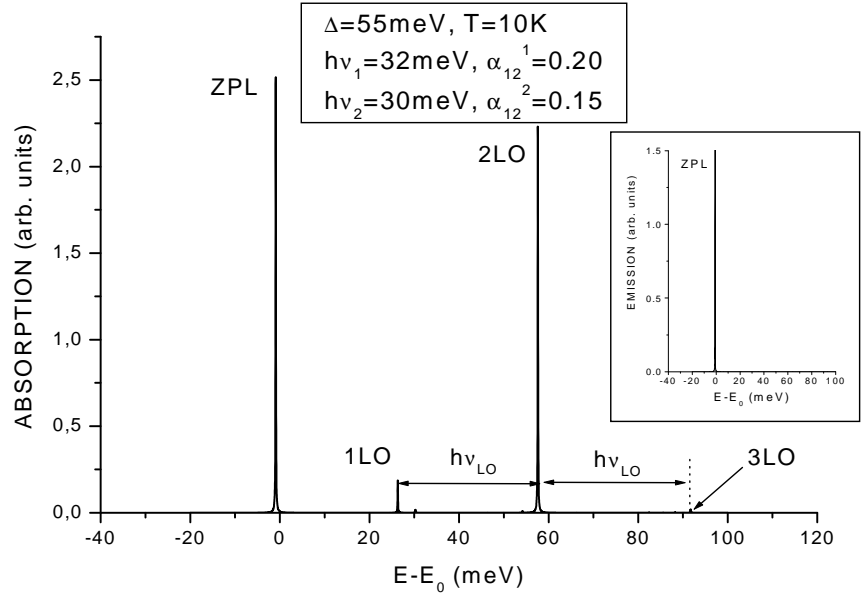}
\caption{ Low temperature absorption spectrum calculated for an
InAs QD with the level spacing of $\Delta =55meV$ assuming two
optical phonon modes (the energies are given on the figure) and
neglecting coupling to acoustic phonons.
The interaction constants used in this calculation are:
$\alpha _{11}^1=0.005$,
$\alpha _{22}^1=0.005$,
$\alpha _{12}^1=0.2$;
$\alpha _{11}^2=0.005$,
$\alpha _{22}^2=0.005$,
$\alpha _{12}^2=0.15$.
Notice that three features above the zero-phonon line are separated
by nearly the same energy, approximately equal to 31meV.
The inset shows the corresponding emission spectrum.}
\end{center}
\label{figure09}
\end{figure}

\begin{figure}[ht]
%Fig.10
\begin{center}
\includegraphics[height=12cm]{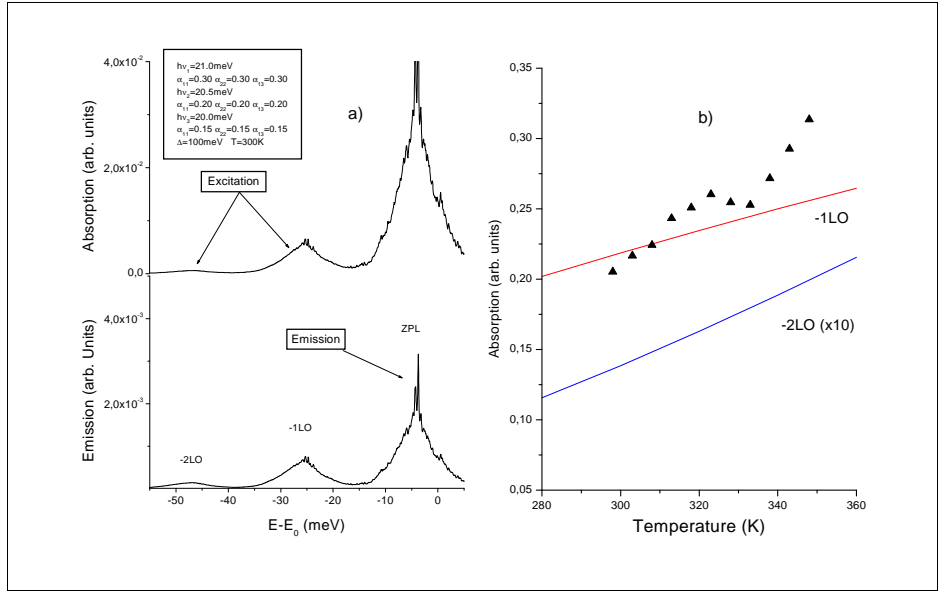}
\caption{ (a) Low-energy part of the absorption and emission
spectra calculated for a hypothetical CdTe QD considering three
optical phonon modes with parameters given on the figure. The
acoustic phonon parameters were taken as for Fig.6. The level
spacing $\Delta =100meV$, temperature 300K. (b) Temperature
dependence of the integrated intensity of two sub-gap bands in the
absorption spectrum (a) (lines) and experimental data of
Ref.\cite{Rak} (points) showing the temperature dependence of the
ASPL peak amplitude.}
\end{center}
\label{figure10}
\end{figure}

\end{document}